\documentclass[useAMS,usenatbib]{mn2e}
\usepackage{journals}

\newcommand{\sub}[1]{_{\mathrm{#1}}}
\newcommand{\ergs}[1]{$10^{#1}$~erg~s$^{-1}$}
\newcommand{\ergscm}[1]{$10^{#1}$~erg~cm$^{-2}$~s$^{-1}$}

\usepackage{url}
\usepackage{amssymb}
\usepackage{graphicx}

\title[Bursts from the RB: indications for a slow rotator]{Indications for a slow rotator in the Rapid Burster from its thermonuclear bursting behaviour}
\author[Bagnoli et al.]{T. Bagnoli$^{1,3}$\thanks{E-mail:
t.bagnoli@sron.nl},
J.J.M. in 't Zand$^{1}$,
D.K. Galloway$^{2}$\thanks{Also at School of Physics and School of Mathematical Sciences, Monash University, Clayton, VIC 3800, Australia.}
and A.L. Watts$^{3}$
\\
$^{1}$SRON Netherlands Institute for Space Research,
Sorbonnelaan 2, 3584 CA Utrecht, The Netherlands\\
$^{2}$Monash Centre for Astrophysics (MoCA), Monash University,
Clayton, VIC 3800, Australia\\
$^{3}$Astronomical Institute ``Anton Pannekoek'', University of Amsterdam,
Postbus 94249, 1090 GE Amsterdam, The Netherlands
\\
\\
\\
\textup{Accepted 2013 February 18.  Received 2013 February 18; in original form 2012 December 6}
}
\begin{document}

\date{}

\pagerange{\pageref{firstpage}--\pageref{lastpage}} \pubyear{}

\maketitle

\label{firstpage}

\begin{abstract}
We perform time-resolved spectroscopy of all the type I bursts
from the Rapid Burster (MXB 1730$-$335) detected with the Rossi X-ray Timing Explorer.
Type I bursts are detected at high accretion rates,
up to $\simeq 45$\% of the Eddington luminosity.
We find evidence that bursts lacking the canonical cooling in their time-resolved spectra
are, none the less, thermonuclear in nature.
The type I bursting rate keeps increasing with the persistent luminosity,
well above the threshold
at which it is known to abruptly drop in other bursting low-mass X-ray binaries.
The only other known source in which the bursting
rate keeps increasing over such a large range of mass accretion rates
is the 11~Hz pulsar IGR~J17480$-$2446.
This may indicate a similarly slow spin 
for the neutron star in the Rapid Burster.

\end{abstract}

\begin{keywords}
stars: neutron -- X-rays: binaries -- X-rays: bursts -- X-rays: individual: MXB 1730$-$335
\end{keywords}

\section{Introduction}

First discovered by \citet{1976ApJ...207L..95L}, the Rapid Burster (MXB 1730$-$335, hereafter RB)
is a neutron star (NS) low-mass X-ray binary (LMXB) located in the globular cluster Liller 1,
with a recent distance measurement of
$7.9 \pm 0.9$~kpc \citep*{2010MNRAS.402.1729V}.
It is a recurrent transient, which had very regular outbursts
every $\sim 210$~d for most of the time
up to MJD $\sim 51500$,
after which the outburst recurrence time abruptly dropped to $\sim 100$~d \citep{2002A&A...381L..45M}.

The RB is unique, in that it features both type I and type II bursts.
While the former are due to the heating and cooling of the surface after
a thermonuclear flash in the underlying layers
(for a recent review see \citealt{2008ApJS..179..360G}, henceforth G08),
type II bursts are most likely due
to the release of gravitational energy from the inner accretion disc
during sudden accretion events (as first pointed out by \citealt*{1978Natur.271..630H}),
probably linked to magnetic gating of the accretion flow (\citealt{1977ApJ...217..197L};
\citealt{1993ApJ...402..593S}; \citealt{2010MNRAS.406.1208D}).
Consequently, these two categories of bursts can differ greatly in duration, recurrence time,
energy and spectral properties
(for a review, see \citealt*{1993SSRv...62..223L}; hereafter LPT93).

Typically, type I burst recurrence times $t\sub{rec}$ are of the order of hours.
Although some type I bursts have been reported to recur within $\sim 200$~s
(e.g. \citealt{2010ApJ...718..292K}) only one source - IGR J17480$-$2446,
hereafter T5X2 - exhibited that behavior for prolonged series of bursts
lasting for hours, instead of multiple events consisting just of two, three or four bursts.
\citet{2012ApJ...748...82L} (henceforth L12) reported T5X2 to constantly increase in burst rate
as its persistent luminosity rose from 0.1 to 0.5 of the Eddington luminosity $L\sub{Edd}$,
with bursts transitioning to a mHz quasi-periodic oscillation (QPO),
with a period as short as 240~s.

This behaviour is well in agreement with models for thermonuclear bursts.
Theory predicts the bursting rate to keep increasing with the persistent luminosity
almost up to the Eddington limit, when stable burning sets in
(\citealt*{1981ApJ...247..267F}; \citealt*{2007ApJ...665.1311H});
still, the burst rate for the majority of burst sources
is observed to decrease dramatically above a few per cent of the Eddington luminosity
(e.g. \citealt{2003A&A...405.1033C}; Fig.~16 in G08).
Although the basic physics of type I bursts is well understood,
this discrepancy between the expected and observed $t\sub{rec}$
remains an unsolved issue in our understanding of the phenomenon.

The unusual properties of T5X2 probably set it apart from the other bursting sources,
and could possibly explain its extremely regular bursting behavior (L12).
T5X2 is an 11~Hz pulsar with a reported magnetic field 
between $10^8$ and $10^{10}$~G (\citealt{2011A&A...526L...3P}; \citealt{2011ApJ...731L...7M}),
making it the slowest rotating burster known, and possibly one of the high magnetic field bursters as well.
Slow rotation could prevent turbulent mixing from stabilizing burning
at low mass-accretion rates
\citep{2009A&A...502..871K}
and reduces the influence of the Coriolis force on flame spread and confinement \citep{2002ApJ...566.1018S}.
Magnetic confinement, which probably accounts for the presence of burst oscillations in T5X2 \citep{2011ApJ...740L...8C},
could limit burning to a constant portion of the stellar surface,
while the burning area may vary with mass-accretion rate in other sources
\citep{2000AIPC..522..359B}.

As already mentioned, the presence of type II bursts is probably indicative
of a prominent dynamic role of the magnetic field $B$ in the RB.
We therefore seek to investigate whether its type I burst behaviour
shows analogies to T5X2 that could be explained in terms
of a slow spin or of surface magnetic field effects,
related to a magnetic gating mechanism in the accretion flow.

We perform a systematic study of all the type I bursts
from the Rapid Burster detected with the Proportional counter array (PCA)
onboard the \textit{Rossi X-ray Timing Explorer} (\textit{RXTE}),
increasing the number of bursts over that already known
by a factor of two.
We observe bursts over one of the largest ranges in persistent flux,
and interestingly find that the burst rate increases to high values
with the mass accretion rate.
We discuss the implications of this behaviour in our understanding
of the peculiar RB.

Section~\ref{sec:obs} describes the data selection, with particular care
devoted to how we excluded spurious bursts (type II, or type I from a different source).
Section~\ref{sec:ana} explains the analyses, spectral and morphological,
performed on the bursts and the persistent emission.
In Section~\ref{sec:res} we report our results on the burst behaviour as a function
of the persistent luminosity of the RB. Finally, we discuss the implications
of our findings for the burst regimes and the source spin
in Section~\ref{sec:dis}.

\section{Observations}\label{sec:obs}

	\subsection{Dataset}\label{sec:dataset}
The PCA \citep{2006ApJS..163..401J} consists of five co-aligned proportional counter units (PCUs)
that combine to a total effective area of 6000~cm$^2$ at 6~keV, in a 2 to 60~keV bandpass.
The photon energy resolution is 18\% full width at half maximum at 6~keV
and the time resolution is programmable down to 1 $\mu$s.
The field of view (FOV) is circular, with radius $\approx1\degr$ (full width to zero response).

The RB has been extensively monitored with the PCA, from the beginning of the mission
in early 1996 until its shutdown in January 2012.
We collected all PCA data that we could find when the RB was in the FOV
of an active PCU, for a total exposure time of 2.83~Msec
(this is the sum of all so-called good-time intervals,
excluding times when RXTE is slewing, near to an SAA passage, or experiences a large particle rate).

A major issue of the PCA data is that the RB has an angular separation of only $0.56\degr$
to the persistently active burster 4U 1728-34. In fact, most observations (67\%) were aimed at 4U 1728-34.
The RB was the aimpoint for 14\% of the time, so that both sources are in the FOV for 82\% of the time
(including as well 1\% when neither source was at the centre of the FOV).
During the remaining 18\%, the PCA aimpoint was offset so that only the RB is inside the FOV,
decreasing the collimator response by a factor $\simeq 0.4$,
but allowing for spectroscopic measurements unaffected by contributions from other sources.
We note that, due to the transient nature of the source,
only 32\% of the data (or 0.91~Msec) was taken whilst the source was actually active.

	\subsection{Identification of RB type I bursts}\label{sec:identification}
We employed the \textsc{Standard 1} data product, consisting of 0.125-s resolved photon count rates per PCU
without any photon energy resolution, to search for bursts. Our method consists primarily of a computer algorithm
to search for statistically significant upward fluctuations above a steady background on time-scales of 1 to 300 s.
We verified the result of this search by a careful visual inspection of the light curves and the identified bursts.
We additionally applied the criterion that the Rapid Burster should be active,
as may be verified with independent RXTE-ASM or ISS-MAXI measurements, to exhibit bursts.
The computer algorithm was set to accept fluctuations above the background
if they have a significance accumulated over the burst that is at least 10 $\sigma$
and if they are accompanied by a stable pointing of the PCA.
The algorithm is efficient as long as there are time intervals before or after a burst within 300~s
where the background can be measured confidently.
As a result some long type II bursts may be missed that have no flat background between them
and bursts that occur during slews with the RB in the field of view.
Through a visual inspection, however, we made sure that type I bursts with the typical fast-rise exponential-decay profile
were included even if they occurred during a slew.
We find a total of 7261 bursts, including 121 type I bursts and 7140 type II bursts. The latter number is a lower limit.

Type I bursts from 4U 1728$-$34 are easily recognizable by their short durations ($\la 10$~s,
typical of H-poor nuclear fuel) and by their brighter (often Eddington limited) peak fluxes,
bimodally distributed about a mean of $9.2 \times$\ergscm{-8} for the bursts showing
photospheric radius expansion (PRE) and $4.5 \times$\ergscm{-8} for normal bursts
(\citealt{2001MNRAS.321..776F}; \citealt{2003ApJ...590..999G}; G08),
a factor of 6 brighter than most bursts from the Rapid Burster.
For an angular separation of 0.56$^{\rm o}$, this ratio decreases by 40\% when the Rapid Burster is at the centre of the FOV
and there are some type II bursts from the Rapid Burster which come close to half the peak flux of bursts from 4U 1728-34.
Still, these are rare and the bursts can be distinguished by their time profile.

It is also straightforward to exclude most type II bursts.
Long-lasting, flat-topped bursts, preceded and followed by dips in the persistent emission
are clearly recognizable as type II,
as are the intermediate-duration bursts showing multi-peaked decays \citep{1991MNRAS.251....1T}.
The shortest type II bursts, the only ones featuring single, short peaks
and a fast rise, are potentially ambiguous.
They however have very short ($<10$~s) decays and 
recur very fast in series of 8-40 bursts,
making their cumulative energy release incompatible with
a thermonuclear origin (LPT93, see Sec.\ref{sec:dis} for further discussion).
When a type I burst hides in a sequence of such bursts it is easily recognized by its much slower decay.

We nevertheless acknowledge that we could have erroneously misidentified a few type I bursts as type II.
We intend to publish a comprehensive overview of all type I and II bursting behavior of the Rapid Burster elsewhere,
with a detailed description of the burst identifications.
Here we concentrate on the 121 type I X-ray bursts that we identified from the Rapid Burster in the PCA data.
65 have already been reported (\citealt{2001MNRAS.321..776F}; G08).
Of these bursts, some have only been partially observed, 
due to their proximity to the beginning or end of a pointing, and
some are surrounded or even covered by trains of fast-recurring type II bursts.
These bursts are not suitable for analysis,
especially when the persistent emission is very low,
because part of the accretion is then taking place through the type II bursts,
implying that in this case the persistent flux does not even qualitatively reflect
the actual mass-accretion rate on the surface,
and a relation between burst properties and the persistent emission should therefore no longer be expected.
Consequently, we only focus on type I bursts
from observations in which no type II bursts were observed.

\section{Analysis}\label{sec:ana}

We set out to determine a number of key parameters for all bursts.

	\subsection{Burst durations and recurrence times}\label{sec:times}

Using the 1-s resolution light curve,
we modelled the tail of the burst light curve, when the intensity is less than 90\%
of the peak intensity minus the pre-burst level,
with one or - when necessary - two exponential decay functions.
We used these fits to determine the burst duration $t\sub{dur}$,
defined as the time difference between the burst start time (see G08)
and the time when the intensity drops below 10\% of the
peak intensity minus the pre-burst level,
when the burst count rates are still significantly above the noise,
so that variations in the persistent count rate do not affect
our estimate of the burst duration.
Generally speaking, single exponentials provide sufficiently good fits
to shorter bursts
($t\sub{dur} \sim 50$ s),
while two exponentials are necessary for longer bursts ($t\sub{dur} \sim 100$ s).
Although such a definition of the burst duration is not affected
by the varying noise level at different persistent count rates,
the burst emission is still not over at $t\sub{dur}$.
Spectra were therefore extracted up to the time $t\sub{end}$
when the intensity decreases to within 2$\sigma$ of the pre-burst level.
We then used the exponential fits to estimate how much fluence of the burst
may have been contained in the far tail of the light curve (i.e., after $t\sub{end}$).
This factor, below 5\% for most bursts,
was added to the burst fluence and to its uncertainty (see Sec.\ref{sec:bursts}),
to avoid a systematic effect towards less energetic bursts at higher
persistent count rates.

\begin{figure}
 \includegraphics[width=\columnwidth]{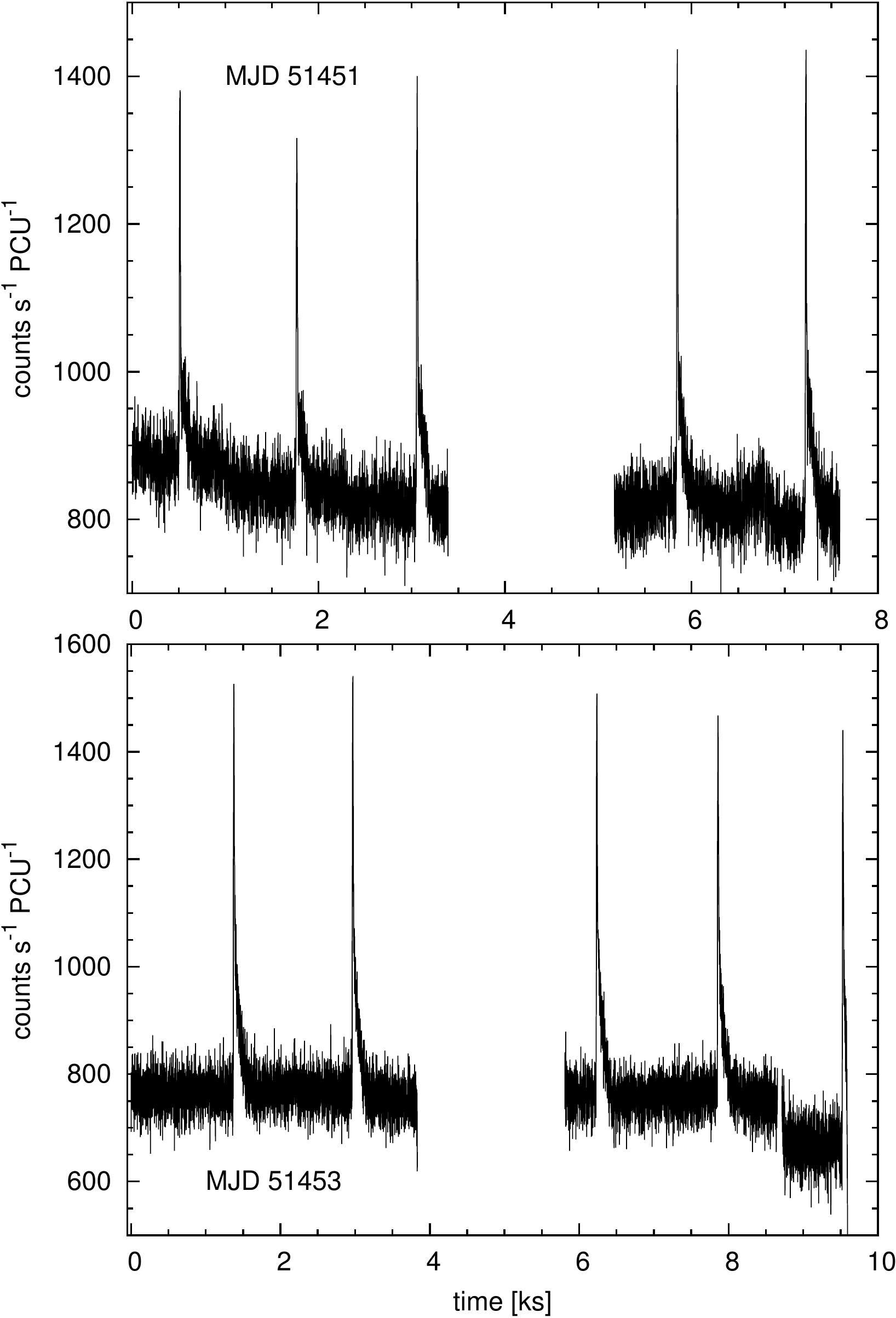}
 \caption{\small{Cooling type I bursts.
Notice the slightly different $x$ and $y$-axis ranges.
The bursts are very regular,
with separations varying by just 10\% in the upper sequence
(as the average $F\sub{pers}$ decreases by roughly the same amount)
and 5\% for the lower sequence.
The burst separations covering the data gaps are close to twice the mean separation of the other bursts,
suggesting another burst occurred in the gap
(see Sec.~\ref{sec:times}).
In the lower panel, the last burst took place in a configuration that excluded 4U 1728-34 from the FOV
and reduced the RB intensity by a factor $\sim 0.4$.
We rescaled the count rate accordingly.
The remaining difference in the persistent emission level
is therefore due to the lack of contribution of 4U 1728-34.}
\label{fig:quintuplet}}
 \end{figure}

The burst recurrence time $t\sub{rec}$ is defined as the time elapsed
between the start time of the previous burst and that of a given burst.
When data gaps occurred, the recurrence time was labelled an upper limit.
In five cases, however, the observed $t\sub{rec}$ was very close
to being an integer multiple ($\sim$2 or 3 times longer)
of the recurrence time observed in the adjacent bursts, which
was seen to be stable over series of up to five bursts,
and a burst occurring with the average value of $t\sub{rec}$
would have fallen in a data gap
(see Fig.~\ref{fig:quintuplet}).
Therefore, we assumed these bursts to be recurring on a regular time-scale
and divided their observed $t\sub{rec}$ by the integer amount of times
that gave a nearly constant burst rate.

	\subsection{Burst energetics}\label{sec:spe}

Throughout our analysis, the particle and cosmic background, as
determined with {\tt ftool pcabackest}, were subtracted from all spectra.
Response matrices were generated with {\tt pcarsp} (v 11.7.1).
All active PCUs were employed, and a correction was applied for
deadtime (although it is always small for the RB).
Low-energy absorption by the interstellar medium was taken into
account using the model of \citet{1983ApJ...270..119M},
with an equivalent hydrogen column density of $N_{\rm H}=1.6\times10^{22}$~cm$^{-2}$
(\citealt*{1995AJ....109.1154F}; \citealt{2000A&A...363..188M}; \citealt{2004A&A...426..979F}).
Following the RXTE Cookbook prescription\footnote{See \url{http://heasarc.nasa.gov/docs/xte/recipes/cook_book.html}},
the analysis was limited to the $3-30$~keV range,
and a systematic error of 0.5\% was added to the results.

To determine the energetics of the bursts and of the persistent emission,
we performed a time-resolved spectroscopic analysis
according to the following procedure.

		\subsubsection{The persistent emission}
First, the spectrum of the non-burst X-ray
emission was extracted from 496~s long \textsc{Standard} 2 data stretches preceding the
burst, to constrain the background signal underlying the burst.
We modelled this spectrum with a generic LMXB model consisting of a disc
black body \citep{1984PASJ...36..741M}, a power law and a Gaussian centred
at 6.4 keV (G08),
that yielded a $\chi^2\sub{red}$ very close to one for all the pre-burst spectra.
In the many cases that the FOV also included 4U~1728$-$34,
the measured flux also included a contribution from that source
and the resulting fit parameters are not a truthful representation of the persistent emission of the RB,
but for assessing the burst energetics this is not crucial
as long as the model is a good description of the non-burst emission.

For those cases where 4U 1728-34 was outside the FOV,
this model also provides a good measurement of the persistent flux $F\sub{pers}$.
Furthermore, even when the FOV contamination was an issue,
a dedicated offset observation (see Sec.~\ref{sec:dataset})
taken no longer than one hour before or after the burst was often available.
We used these uncontaminated spectra to estimate the persistent flux $F\sub{pers}$ in the range 3-25~keV,
resorting to the same three-component model mentioned earlier.
All but six such measurements were taken within 1500~s from the burst onset.
In all cases, we took care to verify the stability of the average count rate
over the interval between the measurement and the burst.
The count rate variation in the RB over the course of an hour was always below 5\%.
All fits were satisfactory, giving $\chi^2\sub{red} \la 1.5$
for 40-50 degrees of freedom.
Note that in the case of short recurrence time bursts,
a single $F\sub{pers}$ measurement may apply to multiple bursts.

For each of these uncontaminated observations,
a bolometric correction $c\sub{bol}$ was calculated
from the portion of the observation excluding 4U~1728$-$34.
We fitted an absorbed Comptonisation model ({\tt compTT} in {\sc XSpec})
to the spectrum extracted for each PCU in the energy range 3--25~keV,
and integrated the best fit model to give the flux over this range.
We then created (in {\sc XSpec}) a ``dummy'' (ideal) response
covering the broader energy range of 0.1--200~keV,
and calculated the integrated, unabsorbed flux over that range.
The bolometric correction for a given PCU was estimated as the ratio of these two fluxes,
and the value adopted for each observation was the mean of the values for the active PCUs,
with the standard deviation of the values over all PCUs adopted as the uncertainty.
The Comptonisation model was chosen over the phenomenological (disk black body plus power law)
because the Comptonisation model has a finite integral for any choice of the input parameters.
In fact, because the persistent spectrum was generally soft,
the electron scattering temperature $kT_e$ was typically a few keV,
and could be well constrained by the spectrum up to 25~keV.
For a few of the observations, the spectrum was harder,
and in some cases the $kT_e$ could not be constrained by the fit.
For most of those observations, the addition of a Gaussian component improved the fit sufficiently
that the $kT_e$ could be constrained;
for a few examples it was possible to constrain the plasma temperature for all but PCU \#3,
and for those observations that spectrum was excluded.

We adopt $F_{\rm pers}=c_{\rm bol}F_{\rm 3-25~keV}$ as the bolometric intensity for the source in each observation.
We also often refer to its ratio to the Eddington flux $F\sub{Edd}$.
To derive the latter, we apply the
distance estimate to the Eddington luminosity

\begin{equation}
 L\sub{Edd} = 3.5 \times 10^{38} \left(\frac{M}{1.4 M_{\odot}}\right)
 \left(\frac{1}{1+X}\right) {\rm ergs~s}^{-1}
\end{equation}
with $M$ the stellar mass
and $X$ the hydrogen fraction (e.g. G08).
This corresponds to an Eddington flux from the disc
$F\sub{Edd} = 2.8\times$\ergscm{-8}
for a ``canonical'' NS with mass $M = 1.4 M_{\odot}$ and radius $R=10$~km,
accreting solar-composition material ($X=0.7$).

		\subsubsection{The bursts}\label{sec:bursts}

The burst data were divided into a number of time bins,
varying in duration to keep the photon count,
and hence the relative error on the derived quantities, approximately constant.
The bin duration was constrained to be at most 1~s during the burst rise,
to be able to track the initially rapidly-varying temperature.
The minimum bin duration possible was constrained
by the time resolution of \textsc{Standard 1} data, at 0.125~s.

Spectra were extracted for each time bin from event files,
the time resolution of which varied between 1 and 125~$\mu$s.
The non-burst emission was not subtracted.
Spectra during the bursts were modelled by the combination of black
body radiation (leaving free temperature and emission area, to account
for the varying burst emission, and multiplied by a model for interstellar absorption
with $N\sub{H} = 1.6 \times 10^{22} \textrm{~cm}^{-2}$)
and the model as found for the non-burst spectrum discussed above,
keeping the parameters of the latter fixed to the pre-burst values.

The assumption underlying this ``standard'' procedure
is that $F\sub{pers}$ is constant during the burst.
The well-known caveat \citep{1986A&A...157L..10V}
is the possibility that the persistent emission actually varies during the burst,
giving rise, for instance, to systematic errors for the radii
that can get very large in the burst tails.
Variations could arise because of radiation effects on the inner accretion disc,
or because of the NS becoming sufficiently hot to contribute to the persistent spectrum.
Worpel, Galloway \& Price (submitted) noted a rise in persistent emission
during radius-expansion bursts which they attribute to radiation drag effects.
We performed a similar analysis, adding a multiplicative factor $f\sub{a}$
by which the persistent emission is multiplied, as a variable parameter in our spectral fits.
However, even for the most luminous burst, the effect appeared negligible,
with $f\sub{a} = 1.06 \pm 0.29$ at the burst peak.
It therefore seems unlikely that this is a major factor for RB bursts,
which are all significantly sub-Eddington.

The black body description of our burst spectra is always
satisfactory. This simple, physically founded model provides a fairly
accurate means to obtain bolometric fluxes and fluences as follows.
The black body fit to each measured spectrum results in the
measurement of two parameters: the temperature, expressed in $kT$, and
normalization $N$, expressed in terms of the square of the radius in
km of a sphere assumed to emit the black body radiation at a distance
of 10 kpc. The values of both these parameters come with a 1-sigma
uncertainty which defines a 68\% confidence margin for a one parameter fit.
The burst bolometric flux is calculated from Stefan-Boltzmann law:

\begin{equation}
F_{\rm bb} = 1.07 \times 10^{-11}~N~\left(\frac{kT}{1\rm{~keV}}\right)^4\mathrm{~erg~cm}^{-2}\mathrm{s}^{-1}. 
\end{equation}

To determine the uncertainty in $F_{\rm bb}$, we sample 10,000 values of
k$T$ and $N$ from Gaussian distributions centered at the fitted values
and with standard deviations equal to the respective 1-sigma
uncertainties, calculate for each sample $F_{\rm bb}$ and from the
10,000 samples calculate the mean and standard deviation. The fluence
is calculated from the sum of all bolometric fluxes times the
integration times of the related spectra and the uncertainty in the
fluence from the root of the quadratic sum of the uncertainty in all
terms. Since the errors in k$T$ and $N$ were treated as if these
parameters are independent, the error in the burst bolometric flux can be
considered as conservative. 

Some uncertainty remains for the
fluence, for the burst radiation in the possibly long tail where it
drowns in the noise. We estimated this tail contribution from the
exponential fits discussed in Sec.~\ref{sec:times}.

		\subsubsection{Determination of $\alpha$ and $\beta$}
The ratio of the integrated bolometric persistent flux to the burst fluence
$\alpha \equiv F\sub{pers} t\sub{rec} /E\sub{b}$
could be precisely determined only for sixteen bursts.
The other bursts suffered from source confusion (see Sec.~\ref{sec:dataset})
or data gaps making $t\sub{rec}$ only an upper limit.
For five additional bursts, however, we could establish the recurrence time
fairly confidently even in the presence of data gaps (see Sec.~\ref{sec:times}),
bringing the total number of bursts with a known value for $\alpha$ to 21,
including four bursts lacking spectral evidence for cooling in their decays (see Sec.~\ref{sec:cool}).
Errors on $F\sub{pers}$ and $E\sub{b}$ are nearly always below 10\%,
and the resulting errors on $\alpha$ are generally below 20\%.

We also looked at the ratio of the burst peak to persistent flux
$\beta \equiv F\sub{peak}/F\sub{pers}$, which we could derive for 57 bursts.
The errors on the peak flux were somewhat large, which propagated
to errors on $\beta$ getting large especially at the lower end of the
$F\sub{pers}$ range.

\section{Results}\label{sec:res}

	\subsection{Lack of cooling}\label{sec:cool}
We fail to find evidence for cooling in six bursts,
meaning the temperature difference $\Delta T = T\sub{peak} - T\sub{dur}$
is consistent with zero within error bars,
with the black body temperature $T\sub{peak}$ measured at the flux peak
and $T\sub{dur}$ at the time $t\sub{dur}$.
They occurred at the beginning of the June-July 1997 outburst,
at the highest level of $F\sub{pers}$ we observed in all data
(see Fig.~\ref{fig:triplet}).
The highest value for which cooling is not observed is $\beta = 0.37 \pm 0.09$,
while the lowest value among bursts with a cooling tail is $\beta = 0.50 \pm 0.10$.
On average $\beta = 0.30 \pm 0.13$ for the non-cooling bursts.
Their $\alpha$ values range from $87\pm13$ to $153\pm26$,
consistent with a thermonuclear origin,
and they have time profiles typical for type I bursts (see Sec.\ref{sec:identification}).
This convinces us that they are not type II bursts,
for which reported values of $\alpha$ range from 0.15 to 4.29
(\citealt{1991MNRAS.251....1T}; \citealt{2000A&A...363..188M}).

	\subsection{Burst behaviour as a function of accretion rate}\label{sec:res:behav}
In Fig.~\ref{fig:sample} all burst parameters are plotted against
the bolometric accretion flux $F\sub{pers}$
and its ratio to the Eddington flux $F\sub{Edd}$.
The RB produces regular type-I bursts over a much wider range
of persistent flux than seen in other systems,
from $(2.7 \pm 0.2)\times$\ergscm{-9}
to $(12.5 \pm 0.3)\times$\ergscm{-9},
or between 10\% and 45\% of $F\sub{Edd}$.

Some trends in burst parameters versus persistent flux
are clearly visible in Fig.~\ref{fig:sample}. The ratio $\beta$
clearly decreases with increasing $F\sub{pers}$,
until it falls below the threshold $\beta = 0.7$
which \citet*{2011ApJ...733L..17L} suggest determines
whether cooling can be detected in PCA data \citep*{2011ApJ...733L..17L}.
This is not only trivially due to the increase in $F\sub{pers}$.
The peak fluxes show a decreasing trend with increasing $F_{\rm pers}$, down from a maximum
of $F\sub{peak} =  (13.6 \pm 2.6) \times$\ergscm{-9}.
On the other hand, the non-cooling bursts at the highest $F\sub{pers}$
show much lower peak fluxes,
averaging $F\sub{peak} \simeq 3.7 \times$\ergscm{-9}.

\begin{figure}
 \includegraphics[width=\columnwidth]{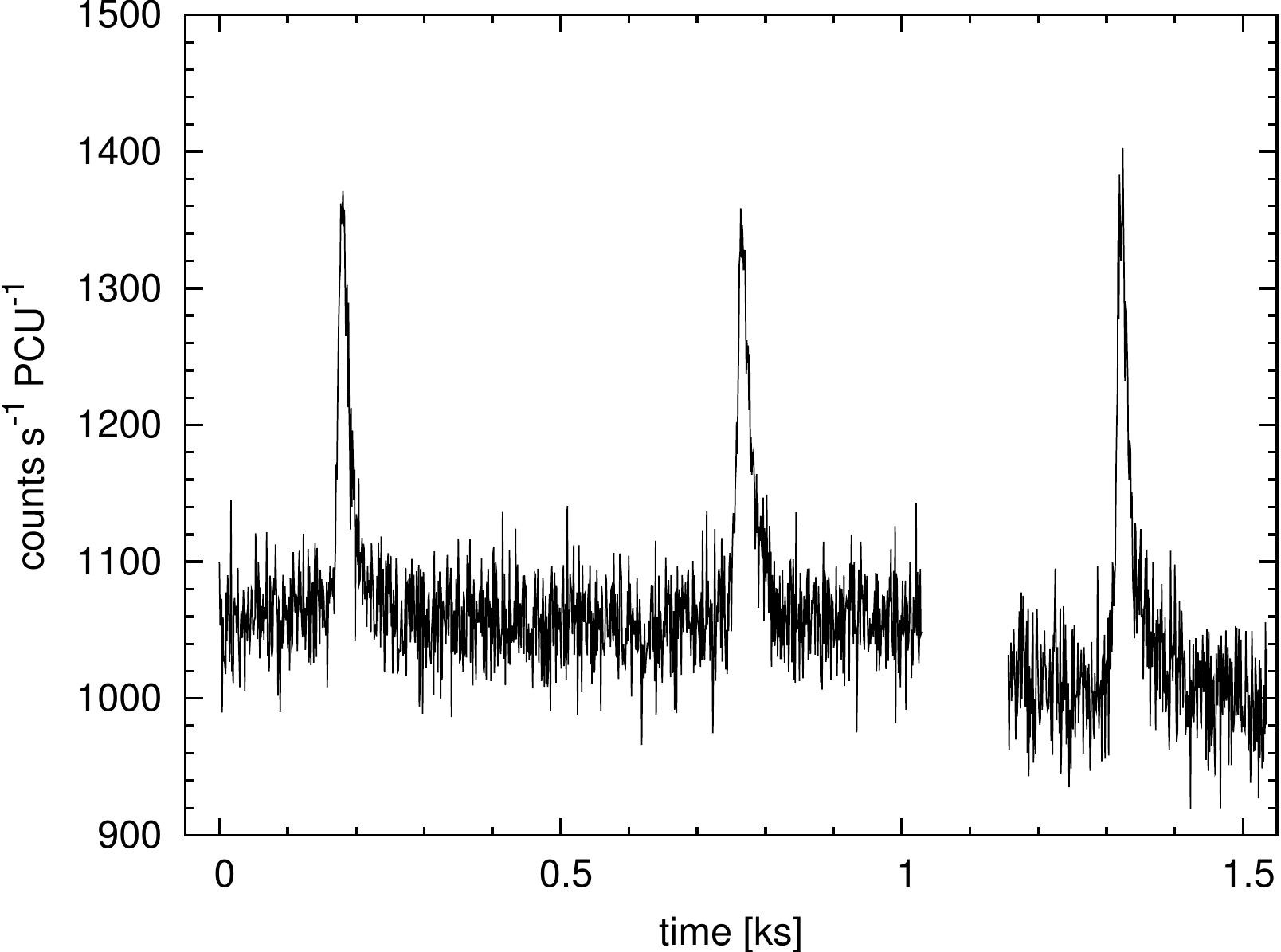}
 \caption{\small{Fast recurring, non-cooling type I bursts.
Concerning the offset in count rate around the last burst, see Fig.~\ref{fig:quintuplet}.}
\label{fig:triplet}}
 \end{figure}

\begin{figure}
\includegraphics[width=\columnwidth]{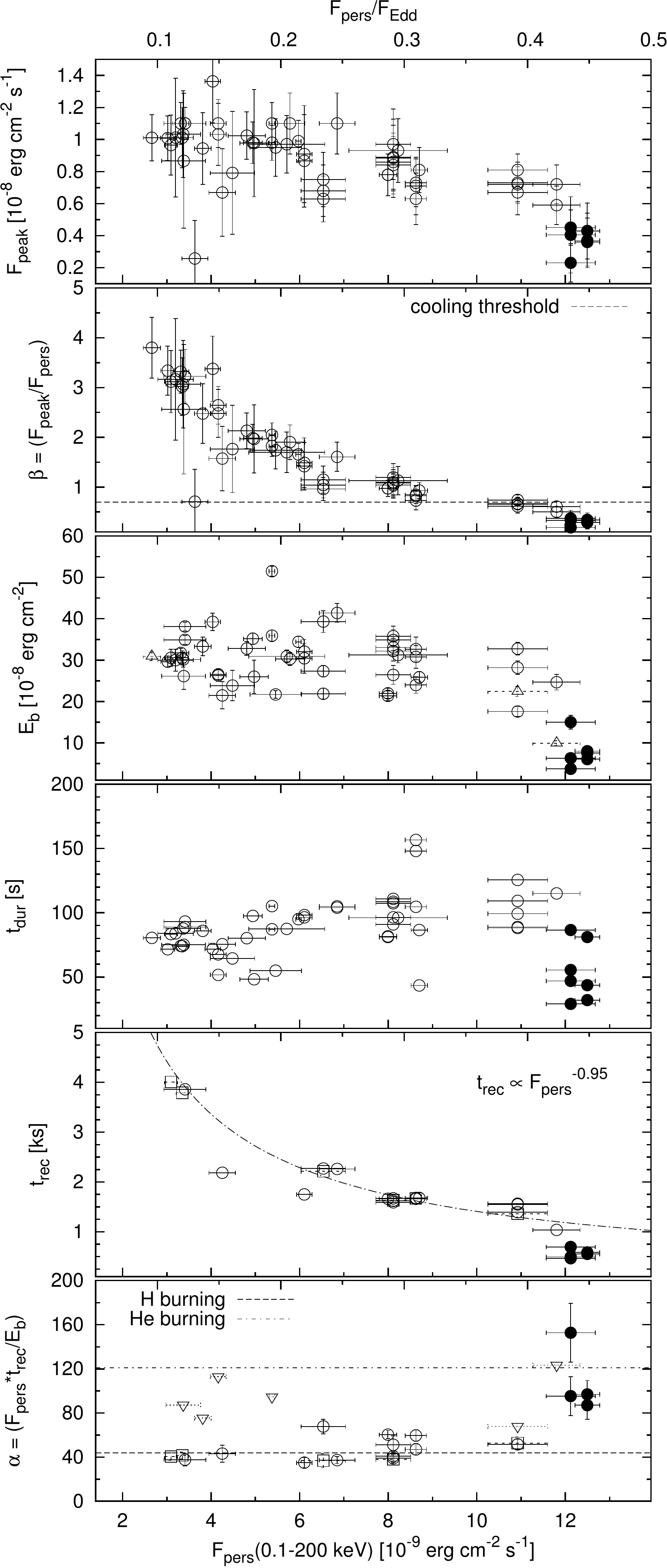}
\caption{\small{Peak flux $F\sub{peak}$, ratio $\beta$, fluence $E\sub{b}$, burst duration $t\sub{dur}$,
burst recurrence time $t\sub{rec}$ and ratio $\alpha$ plotted against the bolometric unabsorbed $F\sub{pers}$ (lower x-axis)
and its ratio to the Eddington flux
(upper x-axis).
Horizontal error bars refer to the lower x-axis.
Empty and filled symbols correspond to cooling and non-cooling bursts respectively,
triangles indicate upper or lower limits,
squares refer to $t\sub{rec}$ values (and the corresponding $\alpha$ values)
that were inferred dividing the observed value 
by an integer to match the ones in the nearby bursts (see Sec.~\ref{sec:times}).
For the ``cooling threshold'' at $\beta = 0.7$, see Sec.~\ref{sec:res:behav}.
The fit to $t\sub{rec}$ is for cooling bursts only.
}
\label{fig:sample}}
\end{figure}

Most importantly, the recurrence time $t\sub{rec}$
falls steadily with increasing $F\sub{pers}$, down to a minimum of 467~s.
Regardless of $F\sub{pers}$, the bursts appear to be very regularly clocked
(see Figs.~\ref{fig:quintuplet} and \ref{fig:triplet}).
A fit to the recurrence times yields a power law with index $-1.81 \pm 0.04$.
The fit however is unacceptably poor, with $\chi^2\sub{red} = 22.1$ for 22 degrees of freedom.
Excluding the non-cooling bursts, this becomes a flatter $-0.95 \pm 0.03$,
with $\chi^2\sub{red} = 3.46$ for 17 degrees of freedom.
The impossibility of fitting a single relation is due
to the steepening of the relation in the non-cooling regime,
as is clear from Fig.~\ref{fig:sample}.
However, too few non-cooling bursts are available
and they span too narrow an $F\sub{pers}$ range
for a fit to this latter regime to be meaningful.

Despite the large spread in values, both $E\sub{b}$ and $t\sub{dur}$
seem to stay relatively constant until the highest persistent fluxes.
A drop then occurs, 
coincidently with the disappearance of cooling from the burst spectra.
One might suspect this to be due to the poorer statistics
in the burst tail at large $F\sub{pers}$ -
a sizeable fraction of the burst fluence might be lost in the noise.
However we defined $t\sub{dur}$ and $E\sub{b}$
exactly to avoid such an effect (see Sec.~\ref{sec:times}).
We therefore believe this drop to be genuine.
The average rise time shows no correlation with the persistent flux
($t\sub{rise} \simeq 7.1$~s, with standard deviation $\sigma \simeq 2.1$~s),
although its average value is different for the cooling and non-cooling bursts
(6.8 and 9.6~s respectively),
perhaps hinting at a late change in bursting regime.
Together with the change in burst duration $t\sub{dur}$,
which on the other hand seems to get shorter at the largest accretion rates,
this makes for bursts becoming somewhat more symmetric in shape.

Excluding the bursts at the highest accretion rates, nearly
constant $\alpha$ values are observed for the cooling bursts, although at lower $F\sub{pers}$
the longer $t\sub{rec}$ means that often only upper limits on $\alpha$ can be measured.
For the cooling bursts for which $\alpha$ is well constrained,
on average $\alpha \simeq 46$ with a standard deviation of $\simeq 10$.
Including the bursts for which we divided $t\sub{rec}$ by an integer value
estimated by comparison with the surrounding bursts
(see Sec.~\ref{sec:times} and Fig.~\ref{fig:quintuplet})
yields an average $\alpha \simeq 44$ with a standard deviation of $\simeq 9$,
reinforcing our inference that our assumption of stable recurrence time is correct.
Only at the high-end of the distribution is an upturn in $\alpha$ visible,
peaking to $\alpha = 153 \pm 26$.
On average, the non-cooling bursts have $\alpha \simeq 98$ with a standard deviation of $\simeq 26$.
It is important to stress that for all the bursts
for which $\alpha$ is well constrained, the value points to a thermonuclear origin.

No obvious explanation is available for the outlier,
the very low $F\sub{peak}$ burst at $F\sub{pers} = 3.6 \pm 0.3\times$\ergscm{-9}.

\section{Discussion and conclusions}\label{sec:dis}

  \subsection{Cooling and burst type identification}

The identification of a burst from the RB as type I or II can be difficult.
Several attempts have been made using broad-band spectroscopy,
but the results are inconclusive at best (\citealt{1984PASJ...36..215K}; \citealt{1988ApJ...324..379S};
\citealt{1991MNRAS.251....1T}; \citealt{1992MNRAS.256..624L}; \citealt{1999MNRAS.307..179G}).
None the less, the presence of cooling in time-resolved spectroscopy
of type I bursts has long been accepted as a hallmark of thermonuclear burning (LPT93),
for type II bursts generally exhibit very little spectral evolution.
This has been used to argue for the occurrence of type II bursts in other sources
(e.g. T5X2; \citealt{2010ATel.3000....1G}).
However several authors (\citealt{2011MNRAS.414.1508M}; \citealt{2011ApJ...730L..23C})
soon pointed out that the $\alpha$ values for T5X2 bursts that did not show a cooling trend
were consistent with a thermonuclear origin.
\citet{2011ApJ...733L..17L} found
the occasional lack of cooling in some bursts from T5X2, Cir X-1, GX 17$+$2 and Cyg X-2
to coincide with a low value for $\beta$ near the outburst peak.
They proved the $\alpha$ values of these bursts
to be consistent with a thermonuclear origin,
and found cooling to disappear from bursts
below a ratio $\beta$ of 0.7.
They attributed the lack of measurable changes in $T\sub{bb}$ to two factors: 
a loss of sensitivity due to the increased luminosity of the persistent emission
(resulting in lower signal to noise);
and a reduction in the column depth required for ignition as the layer temperature rises
(resulting in less mass being burned and the bursts reaching lower peak temperatures).
Values of $\beta$ for non-cooling PCA bursts from Cir X-1, GX 17$+$2 and Cyg X-2
are also below $\beta \simeq 0.7$,
ruling out these sources as type II burst emitters.

Ultimately, $\alpha$
remains the most reliable observable to distinguish type I from type II bursts.
Depending on the fuel composition, thermonuclear burning can release between 1.6 and 4.4~MeV/nucleon
(\citealt{1981ApJS...45..389W}; \citealt{1987ApJ...319..902F}),
while the energy release from accretion on to a ``canonical'' neutron star is about 180~MeV/nucleon,
yielding expected $\alpha$ values between $\sim 40$ and $\sim 110$ for thermonuclear bursts,
about one to two orders of magnitude above what is observed for type II bursts
(\citealt{1991MNRAS.251....1T}; \citealt{2000A&A...363..188M}).
We found compelling evidence that RB bursts at the highest persistent fluxes
have a thermonuclear origin with precise measurements of their $\alpha$ values,
and showed that cooling disappears below a $\beta$ threshold that is roughly consistent
with that inferred by \citet{2011ApJ...733L..17L}.
Although the transition {seems to lie at a somewhat lower value of $\beta$ in the RB,
it should be noted that the energy band they used to calculate $F\sub{pers}$
is narrower than ours ($2-50$~keV versus $1-200$~keV).
Also, \citet{1999MNRAS.307..179G} already noted that in the RB
type II bursts only appear late in the outburst decay,
usually after the persistent luminosity has decreased to  $F\sub{pers} \sim 3 \times$\ergscm{-9}.
For all but one burst we analysed, $F\sub{pers}$ lies above this value
and we confirm that no type II bursts are observable in the lightcurves.
It is therefore unlikely that type II bursting activity
resumes at the highest $F\sub{pers}$ levels,
with $\alpha$ values two orders of magnitude larger than usual
and smooth exponential decays rather than the multi-peaked decays
type II bursts of similar duration are reported to show \citep{1991MNRAS.251....1T}.

The paucity of identified non-cooling bursts is due not only to the relatively short time the source
spends at the highest accretion rates during an outburst,
but also to the drop in outburst peak flux after MJD~51500.
\citet{2002A&A...381L..45M} reported 
a drop in both the outburst peak flux and recurrence time
by roughly by a factor of two after that date.

	\subsection{Bursting regimes}
We showed that the RB keeps emitting type I bursts at an ever
increasing rate, over a range of $F\sub{pers}$ 
from $(2.7\pm 0.2)\times$\ergscm{-9} to $(12.5 \pm 0.3)\times$\ergscm{-9}.
Assuming a distance of 7.9~kpc, this corresponds to $(2.6 - 9.3)\times$\ergs{37},
or $(10-45)$\% $L\sub{Edd}$.
The X-ray flux is however known to be a poor proxy of the real mass accretion rate
\citep{2001ApJ...561..943V}, so that the real range in $\dot{M}$
that the source is spanning remains uncertain.
Accounting for the uncertainty in the distance,
the maximum luminosity at the time of a burst ranges between 35\% and $53\%$ of $L\sub{Edd}$.

That the bursting rate keeps increasing over such a large range in $F\sub{pers}$ is
a striking similarity between the RB and T5X2,
and it sets these two sources apart from all other known bursters
(e.g. \citealt{2003A&A...405.1033C}; Fig.~16 in G08).
The trend in $F\sub{peak}$ is also similar. 
As mentioned, we do not observe a continuous decrease in $E\sub{b}$
and $t\sub{dur}$.
These trends are, however, rather dubious in T5X2 
when looking at the individual bursts, and only become clear switching to
the daily averages of these burst properties (see Fig.~7 and 9 in L12).
A larger number of available bursts in T5X2 is due to a bursting rate that initially grows faster:
$t\sub{rec} \propto F\sub{pers}^{-3}$ up to $F\sub{pers}/F\sub{Edd} \simeq 0.3$,
after which it settles on a flatter $t\sub{rec} \propto F\sub{pers}^{-1}$.
The latter relation seems closer to the one followed by the RB for all but the highest $F\sub{pers}$ levels:
excluding non-cooling bursts, we fit $t\sub{rec} \propto F\sub{pers}^{(-0.95 \pm 0.03)}$.
This is consistent with the empirical relations
$t\sub{rec} \propto F\sub{pers}^{-1.05}$
found in GS 1826-24 \citep{2004ApJ...601..466G}
and $t\sub{rec} \propto F\sub{pers}^{-1.1}$ in the accreting millisecond pulsar
IGR J17511-3057 \citep{2011A&A...529A..68F},
and it is indicative of a fuel with a significant H fraction,
where the proportionality expected is roughly $t\sub{rec} \propto \dot{m}^{-1}$,
while the steeper index initially observed in T5X2 agrees with models for pure He burning
\citep{2000ApJ...544..453C}.
In the RB there seems instead to be a transition from a flatter to a steeper regime,
but the data is too limited to measure the slope of the latter.

All bursts have $\alpha$ values between $35 \pm 5$ and $153 \pm 26$,
with a weighted average $\alpha = 44 \pm 14$.
Values of $\alpha$ for the cooling bursts
are indicative of a fuel composition that is compatible with the solar H abundance,
in line with the observed long burst durations typical of the H burning rp-processes.
The higher $\alpha$ values that are measured at the highest levels of $F\sub{pers}$
are unlikely to be due to a smaller H fraction in the burst fuel,
as this is the opposite of what would be expected for the burst with the
shortest $t\sub{rec}$.
Instead, the sudden rise in $\alpha$
might be due to the onset of semi-stable He burning in between the bursts.
If some fraction of the fuel is burning in between bursts,
this will make overall less material available to burn in a burst,
therefore reducing the fluence and making $\alpha$ grow larger.

This explanation seems to be backed up by theoretical modelling
of the burst time-scales and energetics in this transitional regime.
\citet{2012ApJ...752..150K} studied models of the NS envelope after a superburst.
Bursting is initially quenched because of the hot envelope;
the first bursts reappearing after the superburst
are less energetic, feature a slower rise 
and a faster decay than the bursts before it.
As the envelope cools down further,
$t\sub{rec}$ become longer, $t\sub{rise}$ shortens, $t\sub{dur}$ lengthens
and the bursts become again as energetic as they were before.
We suggest that the same transition could be taking place in the RB
as the mass accretion rate decreases during the outburst decay.
Unfortunately, no observations are available for the time prior to the appearance of the non-cooling bursts
to look for stable burning or the mHz QPOs expected from marginally stable burning
at higher mass accretion rates.

It is also worth pointing out that we are not looking at the so-called short-waiting-time bursts
described by \citet{2010ApJ...718..292K}, which are thought to be due to incomplete burning
of the fuel in the preceding burst, leading to rapidly following, smaller bursts,
in sequences of up to four events.
Within such sequences, the bursts following the first one are on average less bright,
cooler, and less energetic than the initial burst, and their decay profiles
lack the longer decay component from the rp-process,
suggesting they take place in a hydrogen-depleted layer.
These short-waiting-time bursts show $\alpha \simeq 5$
\citep{2007A&A...465..559B}, a value so low that the fuel must be left over from
the preceding burst.
We therefore exclude this incomplete-burning scenario to explain the very short $t\sub{rec}$
we report in the RB, given the stability of the burst properties across multiple events,
the long burst tails indicating the presence of H-burning processes
and the aforementioned $\alpha$ values typical of ordinary type I bursts
and the largest in the sample.

	\subsection{Bursts at the highest $\dot{M}$: indications for a slow spin?}
That the bursting rate keeps increasing over such a large range in $F\sub{pers}$ is
a striking similarity between the RB and T5X2.
Such behaviour, although predicted by theory 
(\citealt{1981ApJ...247..267F}; \citealt{2007ApJ...665.1311H})
has so far been reported in no other source.

When analysing a sample of BeppoSAX Wide Field Camera (WFC) data
on nine X-ray bursters, both transient and persistent sources in the Galactic centre,
\citet{2003A&A...405.1033C} found the burst rate to peak at
$L \simeq (1.4-2.1) \times$\ergs{37}
before dropping by a factor of five.
Above this threshold, all sources also showed significantly shorter bursts,
recurring less regularly.
It is difficult to interpret this change in behaviour.
\citet{2003A&A...405.1033C} suggested that it might be due to the transition from unstable
to stable H burning, so that bursts above the critical luminosity are in a pure He regime
(the so called case 2, \citealt{1981ApJ...247..267F}),
where they take a longer time to build up and have much shorter ($\la 10$~s)
tails due to the absence of prolonged H burning via the rp-processes (LPT93).
He burning has a much stronger $T$ dependence, meaning local conditions set by
perturbations in the layer steadily burning H have a larger importance in setting ignition conditions,
leading to variations in $t\sub{rec}$.
This seems however an unlikely explanation.
Firstly, $\alpha$ values should increase significantly
while switching to the less efficient burning of He,
contrary to the result of G08, who reported $\alpha$ values above the threshold
to stay constant or decrease slightly.
Secondly, after an initial drop, the bursting rate should start again increasing
through the transition from case 2 to case 1 (the He triggered, mixed H/He bursts),
again contrary to what is observed.
Thirdly, case 1 bursts have been observed
at lower mass accretion rates (e.g. in GS~1826$-$24, \citealt{2004ApJ...601..466G}),
suggesting the transition from case 3 to case 2 takes place at lower $\dot{M}$.
Another possibility is that this threshold corresponds to the onset
of semi-stable He burning (e.g. \citealt{2003A&A...411L.487I}),
which starts depleting the layer of nuclear fuel.
The theory of thermonuclear burning however places such a transition at near- or above Eddington
luminosities, inconsistent with the low
luminosity (around 10\% of $L\sub{Edd}$) at which it is observed (\citealt{2003A&A...405.1033C}; G08).
A final possibility involves variation in burning area.
As pointed out by \citet{2000AIPC..522..359B}, the observed flux only reflects the overall mass accretion rate $\dot{M}$,
while the burning regime depends on the mass accretion rate per unit area $\dot{m} \equiv \dot{M}/A\sub{eff}$,
with $A\sub{eff}$ the surface area of the igniting region.
If this area increases with a stronger than linear dependence on $\dot{M}$,
$\dot{m}$ might actually \textit{decrease} with increasing persistent flux.

T5X2, by contrast, is very different and behaves more in line with theoretical expectations.  
L12 reported that T5X2 showed a constant increase in burst rate
as its persistent luminosity rose from 0.1 to 0.5 $L\sub{Edd}$,
with bursts gradually evolving into a mHz QPO (with a period as short as 240~s) until the onset of stable burning.  
They posited that this (up to then) unique match between theory and observations
was due to the very slow spin of T5X2.  Spinning at only 11~Hz, T5X2 rotates an order of magnitude more slowly
than all other bursters whose rotation rate has been measured via burst oscillations.
At 11 Hz, rotation is only minimally dynamically relevant,
whereas at 100 Hz and above its effects are much stronger \citep{2012ARA&A..50..609W}.
Rotation should, for example, affect the burning process (and its stability) 
by triggering turbulent mixing \citep{2009A&A...502..871K}.
This would stabilize He burning at lower accretion rates per unit area than for a non-rotating source,
skewing the theoretical prediction.

Given the relatively strong $B$ field in T5X2,
one might wonder whether it is the magnetic field rather than the rotation rate
that determines the unusual bursting behaviour in this source.
Although the field is weak enough that fuel confinement is unlikely to be effective
\citep{1998ApJ...496..915B},
\citet{2011ApJ...740L...8C} argued that dynamical magnetic effects may act
to confine the spreading flame front after ignition,
leading to stalling and explaining the presence of burst oscillations in this source.
Magnetic channelling should also lead to a reduction in shearing instabilities and rotational mixing
in the surface layers independent of the slow rotation (L12).
Without further study, it has been hard to distinguish these two possibilities for T5X2.
However the RB may provide a clue, since in this source there is no evidence for magnetic channelling
(in the form of persistent accretion-powered pulsations).

If rotation rate is the main factor determining the bursting behaviour in T5X2,
then the similarity exhibited by the RB implies that it too may rotate slowly.
\citet{2012ApJ...752...33P} argued that T5X2 has entered its spin-up phase
in an exceptionally recent time (a few $10^7$~yr),
and that also the total age of the binary is extremely low ($\lesssim 10^8$~yr).
It therefore has not only a slow spin,
but also a relatively high magnetic field (between $10^8$ and $10^{10}$~G,
\citealt{2011A&A...526L...3P}; \citealt{2011ApJ...731L...7M}).
If the RB also rotates slowly, one may speculate that it too is at a similar evolutionary point,
a young LMXB with a strong magnetic field.
Where it differs from T5X2, however, is that it also exhibits Type II bursts.
Interestingly, the only other source known to exhibit type II bursts
is the slow pulsar GRO~J1744$-$28,
which spins at just $2.14$~Hz \citep{1996Natur.381..291F}.
It has a reported magnetic field in the range $(2-20) \times 10^{10}$~G
(\citealp{1996ApJ...465L..31S}; \citealp{1997ApJ...482L.163C}),
intermediate between the weak fields ($B \simeq 10^8-10^9$~G)
thought to reside in LMXBs and the strong fields ($B \simeq 10^{12}-10^{13}$~G)
found in X-ray pulsars.
The $B$ value for GRO J1744-28 is probably large enough to confine
burning to a stably burning portion of the surface,
explaining the lack of type I bursts.

Confirming the spin rate of the RB is unfortunately difficult,
since to date there have been no observations
of either accretion-powered pulsations or burst oscillations from this source. 
A low-significance claim of a 306~Hz modulation in a set of stacked burst spectra
was reported by \citet{2001MNRAS.321..776F}.
However results from stacked burst spectra have in the past proved misleading,
see for example the case of EXO 0748-676: 
stacked burst spectra suggested a spin of 45 Hz \citep{2004ApJ...614L.121V}
whereas the true burst oscillation frequency is now known to be 552~Hz \citep{2010ApJ...711L.148G}.
G08 have also already searched for burst oscillations,
but only for frequencies above 10~Hz.
We are planning a more thorough and dedicated investigation,
to look into the overall timing behaviour of the source,
with particular regard to low-frequency phenomena.
(Bagnoli et al., in prep.).

The lack of persistent pulsations or magnetically-confined burst oscillations
could be explained 
by scattering in an optically thick Comptonizing cloud,
which might degrade the signal and hide the modulation
\citep*{2002ApJ...576L..49T},
although this seems to be an unlikely explanation for at least three other LMXBs
\citep*{2007ApJ...659..580G}.
Alternatively, no pulsations would be observed if the dipole moment of the magnetic field
were very closely aligned with the rotation axis of the NS \citep{1993ApJ...408..179C}.
Indeed this would be supported by models for type II bursts (e.g. \citealt{1993ApJ...402..593S})
that involve unstable disc-magnetosphere interactions
in which matter builds up at the magnetospheric radius
until a critical gas pressure is reached.
This would not be possible in a highly inclined dipole model,
where matter can always stream directly to the star along a field line.
More stringent constraints come by \citet{2011MNRAS.416..893D,2012MNRAS.420..416D},
who argue episodic accretion occurs because of an inner disc radius
hovering around the corotation radius $r\sub{c}$
(the radius at which the Keplerian frequency equals the NS spin).
They notice that the kind of instability they model
would not be possible in case of a strongly misaligned dipole,
where at different longitudes the disc would be truncated inside and outside $r\sub{c}$,
thereby always allowing for some accretion to occur
\citep*{2006ApJ...639..363P}
and preventing the buildup of a mass reservoir to release later in a type II burst.
Although such a strongly aligned configuration might seem unlikely,
especially for relatively young systems,
this is in agreement with the fact that we only know two type II burst sources,
compared to over a hundred type I bursters.
T5X2 could perhaps not show type II bursts
because it is a more strongly misaligned rotator than the RB and the bursting pulsar.

\section*{Acknowledgments}

We wish to thank
Diego Altamirano,
Caroline R. D'Angelo,
J\'{o}el K. Fridriksson,
Michiel van der Klis,
Manu Linares
and Alessandro Patruno 
for the occasional guidance and useful discussions during the preparation of this paper.

 \bibliography{paper}{}
 \bibliographystyle{mn2e}

\label{lastpage}
\end{document}